\documentstyle[twocolumn,aps]{revtex}
\begin{document}
\draft

\twocolumn[\hsize\textwidth\columnwidth\hsize\csname@twocolumnfalse%
\endcsname
\title{\Large{ \bf{\center 
Self-Assembled Triply Periodic Minimal 
Surfaces as moulds for Photonic Band Gap Materials}}}

\author{L. Mart\'{\i}n-Moreno$^*$, F. J. Garc\'{\i}a-Vidal $^{\dagger}$
and A. M. Somoza $^{\ddagger}$}
\address{$^*$ Departamento de F\'{\i}sica de la Materia Condensada,
ICMA-Consejo Superior de Investigaciones Cient\'{\i}ficas,
Universidad de Zaragoza, Zaragoza 50015, Spain.}
\address{$^{\dagger}$ Departamento de F\'{\i}sica Te\'orica de la Materia
Condensada,
Universidad Aut\'onoma de Madrid, Cantoblanco, 28049, Madrid, Spain.}
\address{$^{\ddagger}$ Departamento de F\'{\i}sica.
Universidad de Murcia. Aptdo. 4021, Murcia E-30080,Spain.}

\maketitle

\begin{abstract}
We propose systems with structures defined by self-assembled triply
periodic minimal surfaces (STPMS) 
as candidates for photonic bandgap materials.  
To support our proposal we have calculated the photonic bands for 
different STPMS and we have found that, at least, the double diamond and
gyroid structures present full photonic bandgaps. Given the great
variety of systems which crystalize in these structures, the diversity
of possible materials that form them and the range of lattice constants
they present, the construction of photonic bandgap materials with gaps
in the visible range may be presently within reach.
\end{abstract}


]

\narrowtext
A three-dimensional (3D) photonic bandgap material (PBGM) 
is a periodic dielectric system with an absolute frequency gap 
for electromagnetic (EM) waves\cite{Joanbook}.
The opening of photonic gaps is a delicate balance between both 
refraction index contrast and topology of the underlying lattice.
Since its proposal\cite{Yablo87}\cite{John87} and 
first construction in the microwave range \cite{Yablo91},  
PBGMs have attracted much attention due to their many remarkable
applications \cite{Joan97},  
especially for gaps in the visible range. 
However, material engineering of 3D periodic systems 
structured at the relevant length scale for optical gaps 
(hundreds of $nm$) is facing technological problems \cite{ho}.
Another promising strategy is mesoscopic self-assembly:  
in this way partial bandgaps have been reported in 
3D periodic structures synthesized from submicron colloidal particles 
\cite{Messeguer} \cite{vos}. 
Within this route,  
an alternative idea has apparently been overlooked:  
the use of self-assembled triply periodic minimal surfaces (STPMS)  
as natural moulds to construct optical PBGMs. STPMS 
are present in different systems  
like the periodic phases found in 
diblock-copolymers\cite{Thomas88} and lipid/water systems\cite{Luzzati}. 
Recently, the predicted\cite{Scriven76} periodic phases in 
microemulsions have also been observed\cite{Strey}. 

All the binary and ternary systems mentioned above have 
the common property of being self-organized at a supramolecular level, 
with morphologies mainly dictated by the shape of 
internal interfaces separating different domains.
In AB diblock-copolymers these internal interfaces are 
formed by the chemical bonds between the A and B blocks, 
whereas in lipid/water systems or in microemulsions 
(of for example water and oil) the amphiphilic molecules 
form a monolayer interface that fully separates the 
polar compounds (water and amphiphilic heads) from 
covalent ones (oil and amphiphilic tails). From the 
physical point of view, the dominant factor that 
controls the shape of these internal interfaces is 
minimization of its surface free energy.  
This last condition is essentially equivalent to area-minimisation 
with fixed volume fractions and leads to the formation of constant 
mean curvature interfaces. When the different domains have approximately 
equal volume fractions, the mean curvature is zero. This basically 
explains why and when zero mean curvature 
interfaces are present in these systems. These kind of surfaces  
are known by mathematicians as {\it minimal} surfaces and, already
in the last century\cite{math}, it was found that some
minimal surfaces can be joined together forming periodic structures 
in all three spatial directions.  
Subsequentently, it was discovered\cite{Luzzati} that minimal surfaces 
appear in a variety of real systems occurring in biology and 
materials science.

Presently known systems with STPMS have three 
appealing characteristics for an optical PBGM:
i) Their lattice constants
can be tuned by varying their constituents: real systems with STPMS 
have lattice constants ranging from one to hundreds of $nm$.
ii) Their morphologies are bicontinuous with cubic symmetry, which is 
known \cite{Joanbook} to be more appropriate to produce photonic 
bandgaps than the close-packed morphologies resulting in 
self-assembled colloidal crystals. 
iii) Flexibility in the composition, either by the appropriate
choice of the constituents, or by modifications through selective 
chemical reactions, may provide the adequate 
refraction index contrast.

However, although promising, these characteristics are not enough to 
secure the opening of photonic bandgaps: as previously mentioned, 
topology also plays a crucial role. 
So, in order to support our proposal of STPMS as possible PBGM,   
we have studied theoretically 
which architectures, if any, among the most commonly observed 
STPMS,  develop bandgaps. We have performed calculations 
for the structures defined by three different minimal surfaces:
simple cubic (P), gyroid (G) and double diamond (D).  
Our computational scheme has two steps: (i) generation of the different
structures and (ii) calculation of the photonic bands for each one of them.
We have restricted ourselves to systems with two different components 
of equal volume fraction domains, 
characterized by their refraction indexes $n_1$ and $n_2$ (for 
definiteness we take $n_1<n_2$), separated 
by a minimal surface. 

There are several mathematical 
techniques (Weierstrass parameterization, finite element 
methods, ...) \cite{terro} for the generation of triply 
periodic minimal surfaces. Based on physical grounds, it is 
expected that these periodic surfaces could be generated 
via minimisation of a Landau-Ginzburg free energy 
functional \cite{Gozdz96}, result that has been proved 
rigorously \cite{Cordoba}.  We have chosen this last approach, finding 
extremal configurations of the functional by solving numerically 
the Cahn-Hilliard equation\cite{Bray94}.
This equation resembles the time evolution of the local
concentration $ c(\vec{r})$ in the dynamical process of phase
separation in binary mixtures of polymers and alloys. The local 
concentration 
is forced to verify a continuity equation, which guarantees
fixed volume fractions during the full dynamical process. 
As the final result we obtain the stable (or metastable)
time-evolved $c(\vec{r})$
evaluated at a 3D regular grid of $N^3$ points covering the unit cell.
Different triply periodic minimal surfaces, defined as the
interface between both constituents, are obtained for different 
initial configurations of $c(\vec{r})$; known solutions are recovered
by initializing $c(\vec{r})$ near to the expected result.
In Figs. 1a, 1b and 1c we show, respectively, the P, G and D structures 
obtained via this method. 

As regards to the photonic band calculation, we have used a finite 
difference time domain (FDTD) method\cite{Ward}. This method requires a
computational time which scales linearly with the number of mesh 
points inside the unit cell and, therefore, is specially suitable 
for systems with a dielectric constant that changes rapidly in space, 
like the ones we are studying. Basically, in the FDTD method an 
arbitrary initial EM wave with a wavenumber $\vec{k}$
is allowed to evolve both in time and space governed by
a discrete version of Maxwell equations. After a certain run time,
long enough to have good spectral resolution, the frequencies
$\omega(\vec{k})$ of EM eigenmodes in the structure are obtained by
Fourier transforming the time-propagated EM fields.    
For each structure we have performed calculations for values of $N$
ranging from $N=14$ to $N=40$. We have checked that the results obtained
converge rapidly with $N$, which gives us confidence on the quality
and consistency of both structural and photonic calculations. 

It is worth pointing out that, although we are thinking on applications 
in the optical regime, our results are valid for any lattice constant 
due to the scaling properties of the macroscopic Maxwell 
equations \cite{Joanbook}.
Gaps can be characterised by two quantities: 
the wavelength at the midgap frequency ($\lambda_{0}$) 
in units of the lattice constant ($a$),  
times the smallest refraction index ($n_1$),  
$\lambda_{0} /(n_1 a)$, and the ``figure of merit", $f$, defined 
as the quotient between the frequency width of the gap and the midgap 
frequency. $f$ is, at the same time, 
a measure of the gap, of the reduction of transmission
of EM waves though a finite number of layers, and of the robustness of 
the gap to small deviations from uniformity of the lattice constant 
throughout the crystal.

We have calculated the photonic bands 
as a function of the index of refraction contrast $n = n_2/n_1$,
for each one of the cited 3D periodic structures.
For the P structure we have not found photonic gap 
for any $n$. This result is in keeping with photonic
band calculations for the, different but related, system of
dielectric spheres assembled in a simple 
cubic lattice. Inset of Fig. 1a shows the low-frequency photonic bands
for the P structure for a representative $n=\sqrt{13}$, showing
the lack of gap in the spectrum due to some degeneracies in points 
of high symmetry.
However, and this is the main outcome of our calculations, 
we do find gaps for the G and D structures, provided there is 
a minimum refraction index contrast between the two materials. This can 
be clearly seen in the insets of Figs. 1b and 1c where the low-frequency 
photonic bands for the G and D structures are shown,
for the same representative refraction index contrast $n=\sqrt{13}$. 
Fig. 2a summarizes our results for the G structure, showing the 
dependence of both $\lambda_{0}/(n_1 a)$ and $f$ with $n$. 
Fig. 2b shows the same quantities as Fig. 2a, this time for the D 
structure. Notice that, due to the fact that midgap wavelengths are 
approximately twice as large as lattice constants (an even larger 
for large $n$ or large $n_1$), photonic bandgaps in the optical 
regime are expected for lattice constants of  
$\approx 200-350$ $nm$ or even smaller.  

It is noteworthy that not only there is a minimum value of $n$ for the
appearance of photonic gaps, but also the quality factor $f$ 
increases with $n$ both for the G and D structures, thus making 
large refraction index contrast a desirable feature. The great
flexibility in the manipulation of known systems with STPMS 
may provide routes to obtain the desired refraction index contrast.
For example, in AB diblock
copolymers, the optical properties can be tailored modifying the 
A and B monomers, triblock copolymers providing even more flexibility.
In lyotropic systems and microemulsions their liquid nature turns into
an advantage. One of the liquids may be solidified (through freezing or
polymerization) and the other one extracted. Moreover,
the domains of one of the constituents may be used as a template for
many different chemical reactions. For instance, in this way
cubic structures of silica with remarkable periodicity have been
already obtained\cite{Monnier93}. The reported
lattice constant was about 10 $nm$ but it might be possible to increase
this value, as pore sizes larger than 100 $nm$ have been found in
disordered silica mesostructures\cite{Mcgrath97}.
Extraction of the surfactant and substitution by another material with
a large dielectric constant may lead to a PBGM in the ultraviolet
regime. Even more interestingly, 
materials presenting STPMS have been used as hosts for synthesizing
semiconductors with very large refraction index\cite{Yang96}. 
Based on all these, we believe that the construction of 
3D PBGMs in the ultraviolet and optical ranges based on STPMS could 
be built using present day chemistry.

\noindent {\bf Acknowledgments.}
We acknowledge useful discussions with Prof. R. Strey and Prof. P. Tarazona, 
and Prof. J. B. Pendry and Dr. A.J. Ward for providing us 
with the FDTD code \cite{Ward} prior to its publication. \newline

{\Large {\bf Figure Captions}} \newline

{\bf Figure 1}. Three unit cells 
for the different triply periodic 
minimal surfaces analyzed in this paper: {\bf (a)} simple cubic P, 
{\bf (b)} gyroid G and {\bf (c)} double diamond   
D. To the right of each structure, the corresponding photonic band 
structure is shown, for the representative case in which the two 
materials separated by the minimal surface have a refraction index 
contrast $n=\sqrt{13}$. The wave vector 
varies across the simple cubic Brillouin zone defined by the 
high-symmetry points: G(0,0,0), X(0,0,$\frac{\pi}{a}$),
M(0,$\frac{\pi}{a}$,$\frac{\pi}{a}$) and
R($\frac{\pi}{a}$,$\frac{\pi}{a}$,$\frac{\pi}{a}$), $a$ being the 
lattice constant of the system. EM frequency is in units of 
$\frac{2 \pi c}{n_1 a}$ where $\frac{c}{n_1}$ is the light velocity 
in the lowest refraction index material. Note the appearance of a 
complete photonic bandgap, highlighted in yellow, in the G and D band 
structures. \newline

{\bf Figure 2}. Quality factor, $f$ (full line) and midgap 
wavelength $\lambda_0$ (dashed line) as a function of the 
refraction index contrast for the {\bf (a)} gyroid and {\bf (b)} 
double diamond structures. $f$ is the quotient between the 
gap width and the midgap frequency. $\lambda_0$ is in units of the 
lattice constant $a$ times the lowest refraction index $n_1$. 
For both structures there is a minimum refraction index contrast to open 
a bandgap, this value being $n \approx 2.5$ for the G case and 
$n \approx 2$ for the D one.

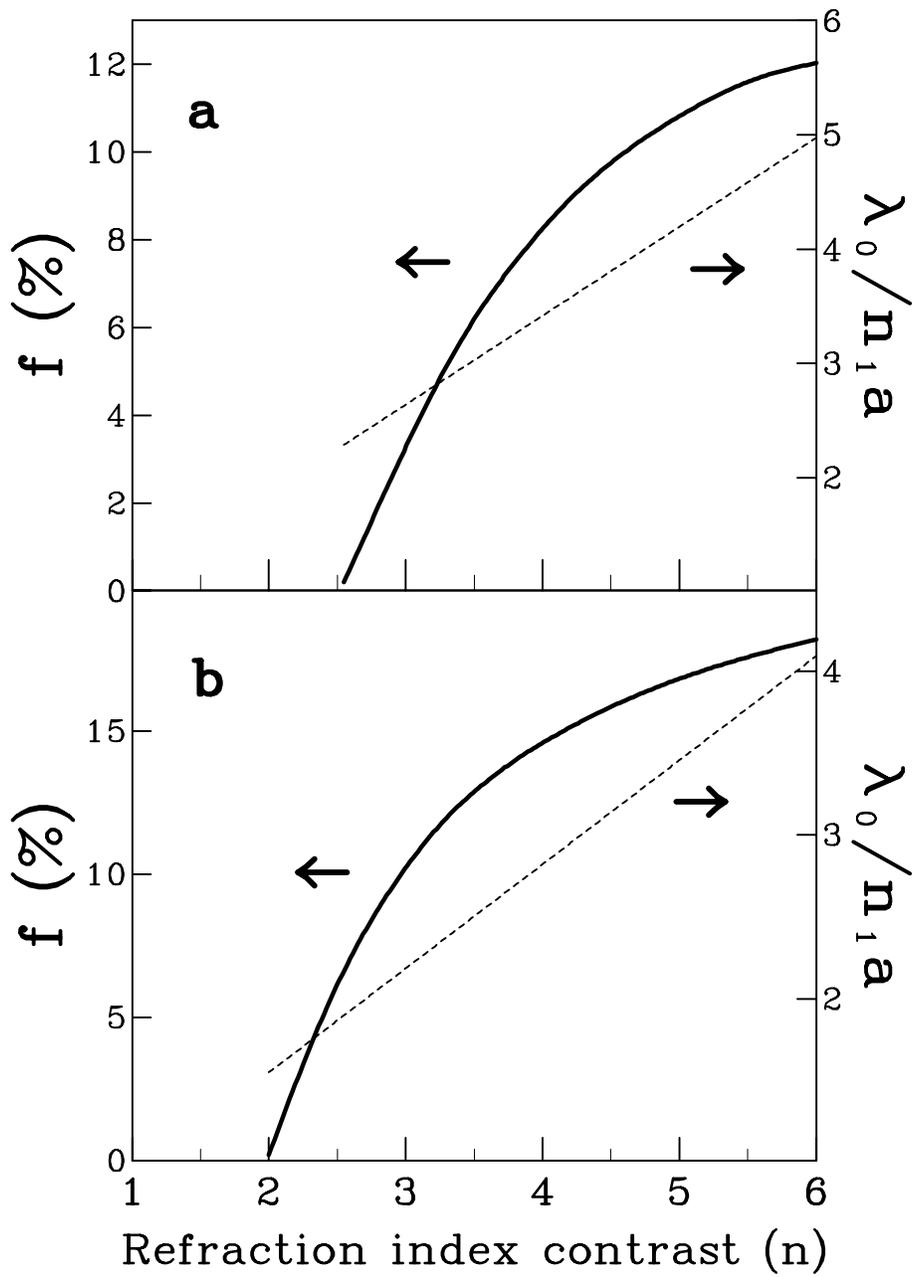
\end{document}